\documentclass[journal]{IEEEtran}

\usepackage{cite}

\ifCLASSINFOpdf
  \usepackage[pdftex]{graphicx}
  \graphicspath{{Imgs/}}
  \DeclareGraphicsExtensions{.pdf,.jpeg,.png}
\else
  \usepackage[dvips]{graphicx}
  \graphicspath{{Imgs/}}
  \DeclareGraphicsExtensions{.eps}
\fi

\usepackage[utf8]{inputenc}
\usepackage[T1]{fontenc} 
\usepackage{amsmath}
\usepackage{amssymb}
\usepackage{mathtools}
\usepackage{amsthm}
\usepackage{mathrsfs}
\usepackage{array}
\usepackage{stfloats}
\usepackage{algpseudocode}
\usepackage{multirow}
\usepackage{xcolor}
\usepackage{tikz}

\newcommand\copyrighttext{%
	\footnotesize \copyright 2023 IEEE. Personal use of this material is permitted. Permission from IEEE must be obtained for all other uses, in any current or future media, including reprinting/republishing this material for advertising or promotional purposes, creating new collective works, for resale or redistribution to servers or lists, or reuse of any copyrighted component of this work in other works.}
\newcommand\copyrightnotice{%
	\begin{tikzpicture}[remember picture,overlay]
		\node[anchor=south,yshift=10pt] at (current page.south) {\fbox{\parbox{\dimexpr\textwidth-\fboxsep-\fboxrule\relax}{\copyrighttext}}};
	\end{tikzpicture}%
}

\newtheorem{thm}{Theorem}
\newtheorem{lem}{Lemma}
\newtheorem{cor}{Corollary}
\newtheorem{defn}{Definition}
\newtheorem{rem}{Remark}

\begin{document}

\title{Efficient Off-Policy Q-Learning for Data-Based Discrete-Time LQR Problems}

\author{Victor~G.~Lopez,
		Mohammad~Alsalti,
        and~Matthias~A.~Müller

\thanks{V. G. Lopez, M. Alsalti and M. A. Müller are with the Leibniz University Hannover, Institute of Automatic Control, 30167 Hannover, Germany (e-mail: \{ lopez, alsalti, mueller \} @irt.uni-hannover.de).}
\thanks{This work has received funding from the European Research Council (ERC) under the European Union’s Horizon 2020 research and innovation programme (grant agreement No 948679).}}

\markboth{IEEE Transactions on Automatic Control}%
{Lopez, Alsalti and Müller: Efficient Off-Policy Q-Learning for Data-Based Discrete-Time LQR Problems}

\maketitle
\thispagestyle{empty}
\pagestyle{empty}
\copyrightnotice

\begin{abstract}
This paper introduces and analyzes an improved Q-learning algorithm for discrete-time linear time-invariant systems. The proposed method does not require any knowledge of the system dynamics, and it enjoys significant efficiency advantages over other data-based optimal control methods in the literature. This algorithm can be fully executed off-line, as it does not require to apply the current estimate of the optimal input to the system as in on-policy algorithms. It is shown that a persistently exciting input, defined from an easily tested matrix rank condition, guarantees the convergence of the algorithm. A data-based method is proposed to design the initial stabilizing feedback gain that the algorithm requires. Robustness of the algorithm in the presence of noisy measurements is analyzed. We compare the proposed algorithm in simulation to different direct and indirect data-based control design methods.
\end{abstract}


\IEEEpeerreviewmaketitle

\section{Introduction}

\IEEEPARstart{F}{or} many control applications, obtaining accurate models of dynamical systems is an impractical task. It can be confidently stated that any attempt to model a real practical system will present at least minor uncertainties. Naturally, the inaccuracies in the models affect the reliability of the controllers designed from them. These circumstances make data-based control an essential area of control theory research.

One approach to address the issue of uncertain system models is regarded as indirect data-based control. This method relies in using data measured from the system to perform a model identification procedure and then using model-based control design. For instance, this approach is used in \cite{DeMaMaReTu:20,DeTuMaRe:19}, where the linear quadratic regulator (LQR) problem is solved by using least square methods to determine the model of the system from noisy data and then using robust control design to approximate the optimal feedback controller.

A different strategy, known as direct data-based control, consists of designing stabilizing controllers from measured data and without an explicit system identification step. Recent developments in data-based analysis of dynamical systems have allowed the design of model-free controllers using this approach. In \cite{Wietal:05}, Willems' et al. showed how to generate any input-output trajectory of a discrete-time linear system using only data collected from the application of a persistently exciting (PE) input. This result is now commonly known as Willems' Fundamental Lemma. To describe their developments, the authors in \cite{Wietal:05} use the framework of behavioral analysis, but their results have been reproduced in the standard state-space representation in \cite{DeTi:20,Waardeetal:20}. Inspired by Willems' lemma, the authors in \cite{Waardeetal2:20} analyze sufficient and necessary conditions for designing controllers using only measured data. From these results, different designs of robust data-based feedback controllers have been proposed, based mostly on the solution of optimization problems with linear matrix inequalities (LMIs) \cite{WaMe:21,PeTe:21,Berbeetal:20,WaCaMe:21}, or using Willems' lemma to design predictive controllers \cite{CoLyDo:19, Berberetal:21}.

In this paper we incorporate the results of Willems' lemma into a reinforcement learning (RL) framework to solve the LQR control problem. RL is a set of algorithms that have been used in recent years as a control design tool for uncertain systems. This method relies on using the previous experience of an agent to improve, and eventually optimize, its performance \cite{SuBa:18}. In \cite{Bert05}, the development of RL algorithms for control of discrete-time systems has been thoroughly studied. 

Q-learning, first proposed by Watkins \cite{WaDa:92}, is an attractive RL method due to its intuitive design and its convergence properties. Moreover, Q-learning is an off-policy algorithm, meaning that the current estimate of the optimal controller is never applied to the system during learning. Instead, the agent learns from the data collected while using an arbitrary PE input. An on-policy version of this method, regarded as Sarsa \cite{SuBa:18}, was used in \cite{BrYdBa:94} to solve the optimal control problem for discrete-time dynamical systems. This iterative method is an on-policy algorithm because it requires the application of the estimated optimal input at every iteration. In any case, the algorithm in \cite{BrYdBa:94} has the important property of being completely model-free. In the more than two decades since its publication, multiple other model-free iterative algorithms have been proposed for continuous-time \cite{MoLeNa:13,JiJi:14,JiJi:12,Luoaetal:14,MoLeJi:15,PaBiJi:21} and for discrete-time systems \cite{KiLeJi:17,Kiumetal:15,MuWaHe:18,FaGeShMe:18}.

This paper makes the following contributions. First, we use Willems' lemma to design a reinforcement learning algorithm that does not require the application of the current estimate of the optimal input to the system, i.e., our method is \emph{off-policy}. This prevents the need to apply a new input and collect new data at every iteration of the algorithm. In contrast, applying the estimated input at every iteration is a requirement in many existing reinforcement learning algorithms \cite{BrYdBa:94,MoLeNa:13,JiJi:14,Kiumetal:15,MuWaHe:18,FaGeShMe:18}. Second, we study the computational complexity of our algorithm and show that it outperforms other data-based methods in the literature that solve the same LQR problem. This computational efficiency comparison is performed in terms of the number of operations required to complete the algorithm, and not in the sense of sample complexity. In particular, although the first off-policy algorithm for discrete-time systems was presented in \cite{KiLeJi:17}, our method requires the computation of fewer parameters and is therefore more efficient. Moreover, we show that our algorithm is significantly more efficient than the interior-point methods that are used to solve optimization problems with LMI constraints as proposed in \cite{WaMe:21,PeTe:21}. Third, we propose to use the method described in \cite{Waardeetal2:20} to determine a deadbeat controller from data, and use this controller as the initial stabilizing policy required to start our algorithm. Finally, we show that the proposed algorithm is robust against small disturbances in the data measurements of the system, providing explicit conditions for the noise that guarantee convergence and stability. To the best of our knowledge, no such bounds have been reported for any policy iteration algorithm in the literature.

This paper is structured as follows. Section \ref{secprel} presents a brief overview of Willems' lemma. The problem formulation and the algorithm design are presented in Section \ref{secalg}. Section \ref{secana} studies the stability and convergence properties of the proposed control method. Section \ref{secdb} shows a solution of the initial-stabilizing-policy problem. In Section \ref{secrob} the robustness properties of our method are studied. Section \ref{seccomp}, compares the computational complexity of the proposed algorithm with existing data-based LQR methods. Simulation studies are presented in Section \ref{secsimul}, where the proposed algorithm is compared against different direct and indirect data-based control methods. Section \ref{secconc} concludes the paper.

\section{Notation and Preliminaries}
\label{secprel}

\subsection{Notation}
\label{secnot}
Throughout this paper, the sample of a signal $x$ taken at time $k$ is denoted by $x_k$. A discrete-time signal is represented as $\{ x_k \}_{k=0}^{N-1} = \{ x_0,\, x_1,\, \ldots,\, x_{N-1} \}$. The Hankel matrix of depth $L$ of a sequence $\{ x_k \}_{k=0}^{N-1}$ is defined as
\begin{equation*}
H_L(x_{[0,N-1]}) = \left[ \begin{array}{cccc} x_0 & x_1 & \cdots & x_{N-L} \\ 
x_1 & x_2 & \cdots & x_{N-L+1} \\ 
\vdots & \vdots & \ddots & \vdots \\ 
x_{L-1} & x_{L} & \cdots & x_{N-1} \end{array} \right].
\end{equation*}

Let $\sigma_{\min}(M)$ denote the minimum singular value of an arbitrary matrix $M$. Moreover, let $P \in \mathbb{R}^{p \times p}$ and $Q \in \mathbb{R}^{q \times q}$ be two symmetric matrices. Then, $\lambda_{\max}(P)$ and $\lambda_{\min}(P)$ are the maximum and minimum eigenvalues of $P$, respectively. The notation $P \succ 0 $ indicates that $P$ is positive definite. Finally, the block diagonal matrix $\text{diag}(P,Q)$ is defined as 
\begin{equation*}
\text{diag} (P, Q) = \left[ \begin{array}{cc} P & 0 \\ 0 & Q \end{array} \right].
\end{equation*}

\subsection{Willems' Fundamental Lemma}
The following is a brief description of Willems' Fundamental Lemma \cite{Wietal:05}. These results were reproduced in the state-space framework in  \cite{Waardeetal:20}. 

Consider a controllable discrete-time linear system
\begin{equation}
x_{k+1}=Ax_k+Bu_k,
\label{linsys}
\end{equation}
where $x_k \in \mathbb{R}^n$ and $u_k \in \mathbb{R}^m$ are the state variable vector and the control input vector, respectively. An input signal $\{ u_k \}_{k=0}^{N-1}$ is said to be persistently exciting (PE) if the conditions in the following definition hold.

\begin{defn}
\label{defpe}
Let $N \geq (m+1)L -1$. An input sequence $\{ u_k \}_{k=0}^{N-1}$ is persistently exciting of order $L$ if the Hankel matrix $H_L(u_{[0,N-1]})$ has full row rank $mL$.
\end{defn}

Definition \ref{defpe} provides a simple method to design a persistently exciting input by means of a rank condition. Notice that this method is easy to satisfy and to verify. Finally, notice that the bound $N \geq (m+1)L -1$ is a necessary requirement for $H_L(u_{[0,N-1]})$ to have full row rank.

Applying a PE input to system (\ref{linsys}) allows us to collect meaningful information about the system dynamics, as stated in the following results.

\begin{lem}[\hspace{1sp}\cite{Wietal:05}]
\label{lemwil}
Let the data $\{ x_k \}_{k=0}^{N-1}$ be collected from a controllable system (\ref{linsys}) when an input $\{ u_k \}_{k=0}^{N-1}$, which is PE of order $n+L$, is applied to it. Then,
\begin{equation}
\text{rank} \left( \left[ \begin{array}{c} H_1(x_{[0,N-L]}) \\ H_L(u_{[0,N-1]}) \end{array} \right] \right) = Lm + n.
\label{wlrank}
\end{equation}
\end{lem}

\begin{thm}[\hspace{1sp}\cite{Wietal:05}]
\label{thwil}
Let the conditions in Lemma \ref{lemwil} hold. Then, $\{ \bar u_k, \bar x_k \}_{k=0}^{L-1}$ is a trajectory of system (\ref{linsys}) if and only if there exists an $\alpha \in \mathbb{R}^{N-L+1}$ such that
\begin{equation*}
\left[ \begin{array}{c} H_L(x_{[0,N-1]}) \\ H_L(u_{[0,N-1]}) \end{array} \right] \alpha = \left[ \begin{array}{c} \bar x \\ \bar u \end{array} \right],
\end{equation*}
where $\bar x = [\bar x_0^\top,\, \ldots,\, \bar x_{L-1}^\top ]^\top$ and $\bar u = [\bar u_0^\top,\, \ldots,\, \bar u_{L-1}^\top ]^\top$.
\end{thm}

Lemma \ref{lemwil} and Theorem \ref{thwil} show that every possible trajectory $\{ \bar u_k, \bar x_k \}_{k=0}^{L-1}$ of the controllable system (\ref{linsys}) can be expressed as a linear combination of time-shifts of a single persistently exciting trajectory collected from (\ref{linsys}). Thus, these results (collectively known as Willems' lemma) allow a data-based representation of a controllable linear system. Willems' lemma has become increasingly popular within the past few years, allowing for data-driven system analysis and controller design. In the following section, we employ this result to design a data-based, off-policy Q-learning algorithm for computing an optimal controller for system (\ref{linsys}).

\section{Off-Policy Q-Learning Algorithm}
\label{secalg}

In this section, we propose a Q-learning algorithm based on Willems' lemma. Different from the policy iteration algorithm proposed in \cite{BrYdBa:94}, we use Willems' lemma to design an off-policy learning method that does not require application of the current estimate of the optimal input to the system. Data from the system dynamics is collected only once during the application of a persistently exciting input. Different from \cite{KiLeJi:17}, our persistence of excitation condition is based on Definition 1 and can be easily tested; moreover, our algorithm requires the computation of fewer parameters (see Section \ref{seccomp}). It is shown that this algorithm converges to the optimal linear quadratic regulator (LQR) solution. Then, we show that our algorithm preserves the stability and convergence properties of its on-policy counterparts.

\subsection{Algorithm design}
Consider the discrete-time linear system in (\ref{linsys}). The pair $(A,B)$ is assumed to be controllable. The objective of the Q-learning algorithm is to determine the solution to the discrete-time LQR problem without any knowledge of the parameters of the model (\ref{linsys}). All that is required is the availability of state and input measurements. 

The LQR control problem is here formulated by defining the one-step cost of using the input $u_k$ at state $x_k$ as
\begin{equation}
c(x_k,u_k) :=  x_k^\top Q x_k + u_k^\top R u_k,
\label{onecost}
\end{equation}
where $Q,R \succ 0$. From (\ref{onecost}), the infinite-horizon cost function is defined as
\begin{equation}
J(x,u) := \sum_{k=0}^\infty c(x_k,u_k).
\label{cost}
\end{equation}
The goal now is to determine the control sequence $\{ u_k \}_{k=0}^\infty$ that minimizes (\ref{cost}). It is well known that this solution is given by $u_k=-K^*x_k$, where 
\begin{equation}
K^*=(R+B^\top P B)^{-1}B^\top P A,
\label{lqrk}
\end{equation}
and $P \succ 0$ is the solution of the discrete-time algebraic Riccati equation \cite{PhNaCh:14}
\begin{equation*}
Q+A^\top P A - P - A^\top P B (R+B^\top P B)^{-1} B^\top P A = 0.
\end{equation*}

It is of our interest to evaluate the cost of using a particular control input of the form $u_k=-Kx_k$. Thus, let $V^K(x_k)$ be the value function under policy $K$, defined as the cost of using the input $u_k=-Kx_k$ from time $k$ onward, that is,
\begin{IEEEeqnarray}{rCl}
V^K(x_k) & := & \sum_{t=k}^\infty \left( x_t^\top Q x_t + u_t^\top R u_t\right) \Big|_{u_t=-Kx_t} \nonumber\\
& = & \sum_{t=k}^\infty x_t^\top \left( Q + K^\top RK \right) x_t.
\label{value}
\end{IEEEeqnarray}
For linear systems (\ref{linsys}), $V^K$ takes the quadratic closed-loop form
\begin{equation}
V^K(x_k) = x_k^\top P^K x_k
\label{valueq}
\end{equation}
for some symmetric matrix $P^K \succ 0$.

Similar to $V^K(x_k)$, define the Q-function under policy $K$, $\mathcal{Q}^K(x_k,u_k)$, as the cost of using an arbitrary input $u_k$ at time $k$, and then using $u=-Kx$ from time $k+1$ onward, i.e.,
\begin{equation}
\mathcal{Q}^K(x_k, u_k) := c(x_k,u_k) + V^K(x_{k+1}).
\label{qfunc}
\end{equation}
Notice that $\mathcal{Q}^K$ and $V^K$ are related as
\begin{equation}
\mathcal{Q}^K(x_k, -Kx_k) \equiv V^K(x_k).
\label{qvsv}
\end{equation}
Substituting (\ref{linsys}), (\ref{onecost}) and (\ref{valueq}) in (\ref{qfunc}) we obtain
\begin{IEEEeqnarray}{rCl}
\mathcal{Q}^K(x_k, u_k) & = & x_k^\top Q x_k + u_k^\top R u_k \nonumber\\
& & + \left( Ax_k + Bu_k \right)^\top P^K \left( Ax_k + Bu_k \right) \nonumber\\
& = & z_k^\top \Theta^K z_k,
\label{qtheta}
\end{IEEEeqnarray}
where $z_k = [x_k^\top,\,\, u_k^\top]^\top$, and
\begin{equation*}
\Theta^K = \left[ \begin{array}{cc} Q+A^\top P^K A & A^\top P^KB \\ B^\top P^KA & R + B^\top P^KB \end{array} \right].
\end{equation*}

Since the Q-function evaluates the cost of using an input $u_k$, solving the minimization problem
\begin{equation}
\mu(x) := \arg \min_{u} \mathcal{Q}^K(x,u) = \arg \min_{u} z^\top \Theta^K z,
\label{minpro}
\end{equation}
where the second equality holds from (\ref{qtheta}), necessarily yields an improved policy $\mu(x)$ in the sense that $V^\mu(x_k) \leq V^K(x_k)$ \cite{BrYdBa:94}. In Section \ref{secana}, we formally show that (\ref{minpro}) is indeed an improved policy. An expression for $\mu$ in (\ref{minpro}) is obtained as
\begin{equation}
\mu_k=-\Theta_{uu}^{-1} \Theta_{ux} x_k,
\label{impu}
\end{equation}
where $\Theta_{uu} \in \mathbb{R}^{m \times m}$ and $\Theta_{ux} \in \mathbb{R}^{m \times n}$ are the submatrices of $\Theta$ defined by
\begin{equation}
\Theta^K \coloneqq \left[ \begin{array}{cc} \Theta_{xx} & \Theta_{ux}^\top \\ \Theta_{ux} & \Theta_{uu} \end{array} \right].
\label{theta}
\end{equation}

Finally, notice from (\ref{qvsv}) that (\ref{qfunc}) allows the recursive expression
\begin{equation}
\mathcal{Q}^K(x_k, u_k) = c(x_k,u_k) + \mathcal{Q}^K(x_{k+1}, -Kx_{k+1}),
\label{qrec}
\end{equation}
which, using (\ref{qtheta}), can be written as
\begin{equation}
z_k^\top \Theta^K z_k = z_k^\top \bar Q z_k + \zeta_{k+1}^\top \Theta^K \zeta_{k+1},
\label{qth}
\end{equation}
with $\bar Q = \text{diag}(Q,R)$ and $\zeta_{k+1}= [x_{k+1}^\top \quad -(Kx_{k+1})^\top]^\top$. Equations (\ref{impu}) and (\ref{qth}) can now be used to construct an off-policy iterative algorithm that solves the LQR problem.

Consider Algorithm 1 below. Here, $\Theta^{i+1}$ is first obtained by solving the set of equations (\ref{qthi}). In Section \ref{secana}, we show that collecting PE data guarantees that (\ref{qthi}) has a unique solution. Then, the feedback gain matrix $K^{i+1}$ is updated as in (\ref{impui}), after partitioning the matrix $\Theta^{i+1}$ in the form of (\ref{theta}). The new matrix $K^{i+1}$ can then be used again in (\ref{qthi}) to start a new iteration. Notice that the input $u_k=-K^ix_k$ is never applied to the system, and (\ref{qthi}) is defined at every iteration using the data collected once in Step 2 of the algorithm. At each iteration, (\ref{qthi}) differs only in the matrix $K^i$ that multiplies the data $x_{k+1}$ in $\zeta_{i,k+1}$. We note that an initial stabilizing feedback gain $K^0$ is required in Algorithm 1; in Section \ref{secdb}, we discuss how such a controller can be computed from data.

\begin{figure}[h]
\hrule
{\bf Algorithm 1: Off-Policy Q-Learning}
{\hrule \small
\begin{algorithmic}[1]
\Procedure{}{}
\State Collect $N\geq (n+1)m+n$ samples of data $\{ x_k, u_k \}_{k=0}^{N-1}$ by applying a PE input of order $n+1$ to the system.
\State Let $\eta = n+m$. Select $\eta$ time instances $\{ k_1,\ldots, k_\eta \}$ such that the vectors $z_{k_j} = [x_{k_j}^\top \,\,\, u_{k_j}^\top ]^\top$, $j=1,\ldots,\eta$, constructed from the collected data, are linearly independent.
\State Let $i=0$ and initialize a stabilizing feedback matrix $K^i$.
\State Using the vectors $z_{k}$ in Step 3, define the set of $\eta$ equations 
\begin{equation}
	z_{k_j}^\top \Theta^{i+1} z_{k_j} = z_{k_j}^\top \bar Q z_{k_j} + \zeta_{i,k_j+1}^\top \Theta^{i+1} \zeta_{i,k_j+1},
	\label{qthi}
\end{equation}
where $\zeta_{i,k+1} = [x_{k+1}^\top \,\,\, -(K^i x_{k+1})^\top ]^\top$. Solve the set of equations for $\Theta^{i+1}$.
\State Update the feedback policy matrix as
\begin{equation}
K^{i+1}=(\Theta_{uu}^{i+1})^{-1} \Theta_{ux}^{i+1}.
\label{impui}
\end{equation}
\State If $\| K^{i+1} - K^i \| > \varepsilon$ for some $\varepsilon > 0$, let $i=i+1$ and go to Step 5. Otherwise, stop. 
\EndProcedure
\hrule
\end{algorithmic}
}
\end{figure}

In the following, we introduce useful lemmas that are then used in Section \ref{secana} to show the convergence and stability properties of Algorithm 1.

\subsection{Basic results}
The analysis performed in this paper makes use of the following preliminary results.
Define the matrix $\Phi_i \in \mathbb{R}^{(n+m) \times (n+m)}$ as
\begin{equation}
\Phi_i = \left[ \begin{array}{cc}
    A & B  \\
    -K^iA & -K^iB 
\end{array} \right].
\label{phi}
\end{equation}

\begin{lem}
\label{lemphi}
The matrix $\Phi_i$ is stable, i.e., has all its eigenvalues strictly inside the unit circle, if and only if the matrix $A-BK^i$ is also stable.
\end{lem}
\begin{proof}
Consider the transformation matrix
\begin{equation*}
	T_i = \left[ \begin{array}{cc}  I_n & 0 \\ -K^i & I_m \end{array} \right] \Rightarrow T_i^{-1} = \left[ \begin{array}{cc}  I_{n} & 0 \\ K^i & I_m \end{array} \right],
\end{equation*}
where $I_p$, $p=\{ n,m \}$, is a $p \times p$ identity matrix and $0$ is a zero matrix of appropriate dimensions. The result follows by noticing that
\begin{equation*}
T_i^{-1} \Phi_i T_i = \left[ \begin{array}{cc}  A-BK^i & B \\ 0 & 0 \end{array} \right]
\end{equation*}
is a block-triangular matrix with diagonal blocks $A-BK^i$ and $0$.
\end{proof}

The matrix $\Phi_i$ is useful in analyzing Algorithm 1 because we can write $\zeta_{i,k+1}$ as
\begin{equation}
\zeta_{i,k+1} = \left[ \begin{array}{c} Ax_k+Bu_k  \\ -K^iAx_k-K^iBu_k 
\end{array} \right] = \Phi_i z_k.
\label{zetaph}
\end{equation}
Notice that, since Algorithm 1 uses a PE input of order $n+1$ in Step 2, by Lemma~\ref{lemwil} (with $L=1$) we have that
\begin{equation}
	\text{rank} \left( \left[ \begin{array}{c} H_1(x_{[0,N-1]}) \\ H_1(u_{[0,N-1]}) \end{array} \right] \right) = n+m.
	\label{rankz}
\end{equation}
Therefore, we can always find a set of $n+m$ vectors $z_{k} = [x_{k}^\top \,\,\, u_{k}^\top ]^\top$ that are linearly independent. This implies that the set of equations (\ref{qthi}) is equivalent to the matrix equation
\begin{equation}
\Theta^{i+1} = \bar Q +  \Phi_i^\top \Theta^{i+1} \Phi_i.
\label{thmat}
\end{equation}

The following corollary follows from Lemma \ref{lemphi}.

\begin{cor}
\label{corsol}
If $K^i$ is stabilizing, then, for any $\bar Q \succ 0$, there exists a unique positive definite solution $\Theta^{i+1}$ of (\ref{thmat}).
\end{cor}
\begin{proof}
From Lemma \ref{lemphi}, since $A-BK^i$ is stable then so is $\Phi_i$. Therefore, (\ref{thmat}) is a discrete-time Lyapunov equation that is known to have a unique solution $\Theta^{i+1} \succ 0$. 
\end{proof}

The following lemma shows a useful expression for $\Theta^{i+1}$.

\begin{lem}
\label{lemide1}
Let $\Theta^{i+1}$ solve (\ref{qthi}) for any stabilizing matrix $K^i$. Then, the following identity holds,
\begin{equation}
\Theta^{i+1} = \sum_{j=0}^\infty (\Phi_i^j)^\top \bar Q \Phi_i^j,
\label{thide1}
\end{equation}
where $\Phi_i^j$ is the $j$-th power of $\Phi_i$.
\end{lem}
\begin{proof}
From (\ref{thmat}), $\bar Q = \Theta^{i+1} - \Phi_i^\top \Theta^{i+1} \Phi_i$. The result is obtained by substituting in the right-hand side of (\ref{thide1}). 
\end{proof}

\begin{cor}
\label{corcose}
The infinite series $\sum_{j=0}^\infty (\Phi_i^j)^\top \Phi_i^j$ converges. Let $\gamma_{1,i} > 0$ be a scalar such that $\gamma_{1,i} = \textup{tr} \left( \sum_{j=0}^\infty (\Phi_i^j)^\top \Phi_i^j \right)$. Then, $\sum_{j=0}^\infty \| (\Phi_i^j) \|^2 \leq \gamma_{1,i}$.
\end{cor}
\begin{proof}
From Corollary \ref{corsol}, if $K^i$ is stabilizing, then (\ref{thmat}) has a solution $\Theta^{i+1}$ for any matrix $\bar Q \succ 0$. Selecting $\bar Q = I$ in (\ref{thide1}) shows the first part of the corollary. The second part is shown by following the relations
\begin{IEEEeqnarray*}{rCl}
\sum_{j=0}^\infty \| \Phi_i^j \|^2 & = & \sum_{j=0}^\infty \lambda_{\max} \left((\Phi_i^j)^\top \Phi_i^j \right) \\
& \leq & \sum_{j=0}^\infty \text{tr} [(\Phi_i^j)^\top \Phi_i^j] \\
& = & \text{tr} \left( \sum_{j=0}^\infty (\Phi_i^j)^\top \Phi_i^j \right) = \gamma_{1,i}
\end{IEEEeqnarray*}
where the inequality holds because $(\Phi_i^j)^\top \Phi_i^j$ is a positive semidefinite matrix. 
\end{proof}

When comparing the solutions $\Theta^i$ of Algorithm 1 at different iterations, we make use of the following identities. Define the matrix $\Psi_i \in \mathbb{R}^{(n+m) \times (n+m)}$ as
\begin{equation}
\Psi_i = \left[ \begin{array}{cc}  0 & 0 \\ (K^i-K^{i-1})A & (K^i-K^{i-1})B \end{array} \right].
\label{psi}
\end{equation}

\begin{lem}
\label{lemide2}
Let $\Theta^i$ and $\Theta^{i+1}$ be the solutions of (\ref{qthi}) at two consecutive iterations of Algorithm 1. The following identity holds,
\begin{equation}
\Theta^i - \Theta^{i+1} = \Phi_i^\top (\Theta^i - \Theta^{i+1}) \Phi_i + \Psi_i^\top \Theta^i \Psi_i.
\label{thide2}
\end{equation}
\end{lem}
\begin{proof}
From (\ref{thmat}), we obtain $\Theta^i - \Theta^{i+1} = \Phi_{i-1}^\top \Theta^i \Phi_{i-1} - \Phi_i^\top \Theta^{i+1} \Phi_i$. Clearly, (\ref{thide2}) holds if $\Phi_{i-1}^\top \Theta^i \Phi_{i-1} = \Phi_i^\top \Theta^i \Phi_i + \Psi_i^\top \Theta^i \Psi_i$. We now prove this relationship. From the definitions (\ref{phi}) and (\ref{psi}), notice that $\Phi_{i-1}=\Phi_i + \Psi_i$. Therefore,
\begin{IEEEeqnarray*}{rCl}
\Phi_{i-1}^\top \Theta^i \Phi_{i-1} & = & (\Phi_i + \Psi_i)^\top \Theta^i (\Phi_i + \Psi_i) \\
& = & \Phi_i^\top \Theta^i \Phi_i + \Psi_i^\top \Theta^i \Psi_i + \Phi_i^\top \Theta^i \Psi_i + \Psi_i^\top \Theta^i \Phi_i.
\end{IEEEeqnarray*}
Moreover, writing $\Theta^i$ as in (\ref{theta}) we can express
\begin{multline*}
\Phi_i^\top \Theta^i \Psi_i = \left[ \begin{array}{cc}  A & B \\ -K^iA & -K^iB \end{array} \right]^\top \left[ \begin{array}{cc}  \Theta_{xx}^i & (\Theta_{ux}^{i})^\top \\ \Theta_{ux}^i & \Theta_{uu}^i \end{array} \right] \\
\times \left[ \begin{array}{cc}  0 & 0 \\ (K^i-K^{i-1})A & (K^i-K^{i-1})B \end{array} \right].
\end{multline*}
Substituting $K^i=(\Theta_{uu}^i)^{-1}\Theta_{ux}^i$ as in (\ref{impui}) in this expression, it is easy to show that $\Phi_i^\top \Theta^i \Psi_i = 0$. Therefore, $\Phi_{i-1}^\top \Theta^i \Phi_{i-1} = \Phi_i^\top \Theta^i \Phi_i + \Psi_i^\top \Theta^i \Psi_i$ completing the proof.
\end{proof}

\begin{cor}
\label{coride3}
Let $\Theta^i$ and $\Theta^{i+1}$ be the solutions of (\ref{qthi}) at two consecutive iterations of Algorithm 1. Then, the following identity holds,
\begin{equation}
\Theta^i - \Theta^{i+1} = \sum_{j=0}^\infty (\Phi_i^j)^\top \Psi_i^\top \Theta^i \Psi_i \Phi_i^j.
\label{thide3}
\end{equation}
\end{cor}
\begin{proof}
Follows from substituting  $\Psi_i^\top \Theta^i \Psi_i = \Theta^i - \Theta^{i+1} - \Phi_i^\top (\Theta^i - \Theta^{i+1}) \Phi_i$ on the right-hand side of (\ref{thide3}).
\end{proof}

We are now ready to study the properties of Algorithm 1.

\section{Algorithm Analysis}
\label{secana}

In this section, we analyze Algorithm 1 to show that the known properties of existing Q-learning methods are maintained if the data $\{ x_k, u_k \}_{k=0}^N$ is obtained from a PE input $u_k$ as in Definition \ref{defpe}. In particular, we show that 1) $\Theta^{i+1}$ can be uniquely determined at every iteration, 2) $K^{i+1}$ generates a stabilizing control policy at every iteration, and 3) Algorithm~1 converges in the limit to the solution of the LQR problem.   

First, we provide a method to solve the set of equations (\ref{qthi}) and show that $\Theta^{i+1}$ can be uniquely determined at every iteration. 
Since the PE condition in Step 2 of the algorithm implies (\ref{rankz}), the collected data provides a set of $\eta = n + m$ linearly independent vectors $z_{k_j}$ as required in Step 3. Using these vectors, define the matrix
\begin{equation}
	Z := \left[ \begin{array}{cccc} z_{k_1} & z_{k_2} & \cdots & z_{k_\eta} \end{array} \right].
	\label{matrixz}
\end{equation}
Thus, $Z \in \mathbb{R}^{\eta \times \eta}$ is a nonsingular matrix. Moreover, define
\begin{equation}
	Y_i := \left[ \begin{array}{cccc} \zeta_{i,k_1+1} & \zeta_{i,k_2+1} & \cdots & \zeta_{i,k_\eta+1} \end{array} \right]
	\label{matrixy}
\end{equation}
with $\zeta_{i,k+1} = [x_{k+1}^\top \,\,\, -(K^i x_{k+1})^\top ]^\top$. Consider now the matrix equation (\ref{thmat}), which is equivalent to the set of equations (\ref{qthi}) (see the discussion above (\ref{thmat})). Pre-multiplying by $Z^\top$ and post-multiplying by $Z$, we obtain
\begin{equation}
	Z^\top \Theta^{i+1} Z = Z^\top \bar Q Z +  Z^\top \Phi_i^\top \Theta^{i+1} \Phi_i Z.
	\label{thmatz}
\end{equation}
Since $Z$ is nonsingular, then a unique solution $\Theta^{i+1}$ of (\ref{thmat}) implies the same unique solution for (\ref{thmatz}).

Now, (\ref{thmatz}) is still a model-based equation due to $\Phi_i$. However, using (\ref{zetaph}) we can write $Y_i$ in (\ref{matrixy}) as $Y_i = \Phi_i Z$. Substituting in (\ref{thmatz}), we get the data-based equation
\begin{equation}
	Z^\top \Theta^{i+1} Z = Z^\top \bar Q Z +  Y_i^\top \Theta^{i+1} Y_i.
	\label{thmatzy}
\end{equation}
Equation (\ref{thmatzy}) is a special case of the generalized Sylvester equation \cite{SaCha:20} and algorithms to solve it efficiently are well known. To summarize, the set of $\eta$ equations in Step 5 of Algorithm 1 can be solved by solving (\ref{thmatzy}). In the following theorem, we show that the procedure in Algorithm 1 guarantees the existence of the solution $\Theta^{i+1}$.

\begin{thm}
	\label{thcons}
	Consider Algorithm 1. Let the matrices $Z$ and $Y_i$ be defined as in (\ref{matrixz}) and (\ref{matrixy}), respectively, using data $\{ x_k,u_k \}_{k=0}^{N-1}$ collected in Step 2 of Algorithm 1. Moreover, assume that matrix $K^i$ in (\ref{qthi}) is stabilizing. Then, the solution $\Theta^{i+1}$ of the set of equations (\ref{qthi}) exists and is unique, and can equivalently be computed by solving (\ref{thmatzy}).
\end{thm}
\begin{proof}
	By Corollary \ref{corsol}, a stabilizing matrix $K^i$ implies that the matrix equation (\ref{thmat}) has a unique solution $\Theta^{i+1}$. As described above, a nonsingular matrix $Z$ implies that the solution of (\ref{qthi}) is equivalent to the solution of (\ref{thmatzy}). The proof is completed by the fact that a nonsingular matrix $Z$ exists by Lemma \ref{lemwil}.
\end{proof}

It is now of our interest to show that Algorithm 1 yields stabilizing policies at every iteration.

\begin{thm}
Let the conditions in Theorem 2 hold. Then, the matrix $K^{i+1}$ in (\ref{impui}) is stabilizing at every iteration of Algorithm 1.
\label{thstab}
\end{thm}
\begin{proof}
The proof is performed by induction. We assume that the feedback matrix $K^i$ is stabilizing and study the stabilizing properties of $K^{i+1}$. Notice that Algorithm 1 initiates a stabilizing matrix $K^0$.

Since $K^i$ is stabilizing, from Theorem \ref{thcons} it follows that (\ref{qthi}) has a unique solution at iteration $i+1$, and hence the control update (\ref{impui}) is well defined. Since (\ref{impui}) is the solution of the minimization problem (\ref{minpro}), where $\mathcal{Q}^{K^i}(x,u) = z^\top \Theta^{i+1} z$, then $\zeta_{i+1,k}^\top \Theta^{i+1}\zeta_{i+1,k} \leq z_k^\top \Theta^{i+1} z_k$ for all $u_k$, where $\zeta_{i+1,k} = z_k |_{u_k=-K^{i+1}x_k}$. In particular, we notice that $\zeta_{i+1,k+1}^\top \Theta^{i+1}\zeta_{i+1,k+1} \leq \zeta_{i,k+1}^\top \Theta^{i+1}\zeta_{i,k+1}$. From (\ref{qthi}), this means 
\begin{IEEEeqnarray}{rCl}
z_k^\top \Theta^{i+1} z_k & \geq & z_k^\top \bar Q z_k + \zeta_{i+1,k+1}^\top \Theta^{i+1} \zeta_{i+1,k+1} \nonumber \\
& = & z_k^\top \bar Q z_k + z_k^\top \Phi_{i+1}^\top \Theta^{i+1} \Phi_{i+1} z_k
\label{lyapsen}
\end{IEEEeqnarray}
where the last equality follows from (\ref{phi}) - (\ref{zetaph}) with $i$ replaced by $i+1$. Since (\ref{lyapsen}) holds for the linearly independent vectors studied in Theorem \ref{thcons}, then it is true for all $z_k$ and the inequality $\Phi_{i+1}^\top \Theta^{i+1} \Phi_{i+1} - \Theta^{i+1} \preceq - \bar Q \prec 0$ holds. Since $K^i$ is stabilizing by assumption, from Corollary \ref{corsol} we obtain that $\Theta^{i+1} \succ 0$ and, therefore, the preceding inequality shows that all the eigenvalues of $\Phi_{i+1}$ are inside the unit circle by using standard Lyapunov arguments. By Lemma \ref{lemphi}, stability of $\Phi_{i+1}$ implies stability of the closed-loop matrix $A-BK^{i+1}$. 
\end{proof}

Finally, we show that Algorithm 1 converges to the optimal LQR solution. The following relationship follows immediately.

\begin{lem}
\label{lemconv1}
Let the conditions in Theorem 2 hold. Then, at every iteration of Algorithm 1, it holds that $\Theta^{i+1} \preceq \Theta^i$.
\end{lem}
\begin{proof}
Follows from Corollary \ref{coride3}, by noticing that the right-hand side of (\ref{thide3}) is positive semidefinite.
\end{proof}

Now, let $K^*$ be the solution of the discrete-time LQR problem as in (\ref{lqrk}), and let $\Theta^*$ be such that
\begin{equation}
\Theta^* = \bar Q +  \Phi_*^\top \Theta^* \Phi_*
\label{thopt}
\end{equation}
where $\Phi_*$ is as in (\ref{phi}) with $K^i=K^*$. Define also the optimal value $V^* := V^{K^*}$. Then Theorem \ref{thcvg} shows that Algorithm 1 converges such that $\lim_{i \rightarrow \infty} \Theta^i = \Theta^*$ and $\lim_{i \rightarrow \infty} K^i = K^*$.

\begin{thm}
If the conditions of Theorem 2 hold, then at every iteration of Algorithm 1 it holds that $\Theta^* \preceq \Theta^{i+1} \preceq \Theta^i$. Moreover, $\lim_{i \rightarrow \infty} \Theta^i = \Theta^*$.
\label{thcvg}
\end{thm}
\begin{proof}
Similar to Lemma \ref{lemide2}, it can be shown that the identity
\begin{equation}
\Theta^{i+1} - \Theta^* = \Phi_i^\top (\Theta^{i+1} - \Theta^*) \Phi_i + \Psi_{*,i}^\top \Theta^* \Psi_{*,i},
\end{equation}
holds, with $\Psi_{*,i}$ defined as in (\ref{psi}) with $K^{i-1}$ replaced by $K^*$. As in Corollary \ref{coride3}, this implies
\begin{equation}
\Theta^{i+1} - \Theta^* = \sum_{j=0}^\infty (\Phi_i^j)^\top \Psi_{*,i}^\top \Theta^* \Psi_{*,i} \Phi_i^j \succeq 0.
\label{idenopt}
\end{equation}
The result $\Theta^* \preceq \Theta^{i+1} \preceq \Theta^i$ then follows from Lemma \ref{lemconv1}.

To prove that $\lim_{i \rightarrow \infty} \Theta^i = \Theta^*$, we first show that $\Theta^*$ is a fixed point of Algorithm 1. Let $\mathcal{K} = (\Theta^*_{uu})^{-1} \Theta^*_{ux}$, and let $\Theta^{\mathcal{K}}$ solve (\ref{qthi}) for $K^i=\mathcal{K}$. As we have shown, this implies $\Theta^* \preceq \Theta^{\mathcal{K}} \preceq \Theta^*$ and, therefore, $\Theta^{\mathcal{K}} = \Theta^*$. We finally show that this fixed point is unique. Assume that Algorithm 1 reaches a fixed point such that $\Theta^{F} = \bar Q +  \Phi_{F}^\top \Theta^{F} \Phi_{F}$, where $\Phi_F$ is defined as in (\ref{phi}) with $K^F= (\Theta^F_{uu})^{-1} \Theta^F_{ux}$. This implies that (\ref{qrec}) holds for $\mathcal{Q}^{K^F}(x_k, u_k) = z_k^\top \Theta^{F} z_k$. Using (\ref{qvsv}), we get $V^{K^F}(x_k) = c(x_k,-K^F x_k) + V^{K^F}(x_{k+1})$. From (\ref{impu}), $u_k=-K^F x_k$ is the minimizer of (\ref{minpro}) and, then, $V^{K^F}(x_k) = \min_{u_k}[c(x_k,u_k) + V^{K^F}(x_{k+1})]$, which is the Hamilton-Jacobi-Bellman equation associated to the cost function (\ref{cost}), known to have a unique solution $V^{K^F}=V^*$ and, therefore, $\Theta^F=\Theta^*$.
\end{proof}

An attractive characteristic of policy iteration algorithms is their quadratic order of convergence \cite{Hewer:71}. This order of convergence means that the order of the error between the approximated and the optimal gains decreases quadratically at every iteration. This property is studied for Algorithm 1 in the following theorem. The proof follows similar steps as \cite[Theorem~2]{Hewer:71}.

\begin{thm}
Algorithm 1 has quadratic order of convergence. In particular,
\begin{equation}
\| \Theta^{i+1} - \Theta^* \| \leq \gamma \| \Theta^i - \Theta^* \|^2
\label{quconv}
\end{equation}
for some scalar $\gamma > 0$.
\label{thqconv}
\end{thm}
\begin{proof}
By direct substitution, it can be shown that $\Psi_{*,i}$ in (\ref{psi}) can be expressed as
\begin{equation}
\Psi_{*,i} = \bar \Theta_u^* (\Theta^i - \Theta^*) \Phi_i,
\label{psinew}
\end{equation}
where the block matrix $\bar \Theta_u^*$ is
\begin{equation}
\bar \Theta_u^* = \left[ \begin{array}{cc}
    0 & 0 \\
    0 & (\Theta_{uu}^*)^{-1}
\end{array} \right]
\end{equation}
and 
\begin{equation}
\Theta^* = \left[ \begin{array}{cc}
    \Theta_{xx}^* & (\Theta_{ux}^*)^\top \\
    \Theta_{ux}^* & \Theta_{uu}^*
\end{array} \right].
\end{equation}
Substituting (\ref{psinew}) in (\ref{idenopt}), we get
\begin{IEEEeqnarray*}{rCl}
\Theta^{i+1} - \Theta^* & = & \sum_{j=1}^\infty (\Phi_i^j)^\top (\Theta^i - \Theta^*) \bar \Theta_u^* \Theta^* \\
& & \hfill \times \bar \Theta_u^* (\Theta^i - \Theta^*) \Phi_i^j \\
& = & \sum_{j=1}^\infty (\Phi_i^j)^\top (\Theta^i - \Theta^*) \bar \Theta_u^* (\Theta^i - \Theta^*) \Phi_i^j
\end{IEEEeqnarray*}
Computing the norm of both sides of the inequality,
\begin{IEEEeqnarray*}{rCl}
\| \Theta^{i+1} - \Theta^* \| & \leq & \sum_{j=1}^\infty \| (\Phi_i^j) \|^2 \| \bar \Theta_u^* \| \| (\Theta^i - \Theta^*) \|^2 \\
& \leq & \gamma_{1,i} \gamma_2 \| (\Theta^i - \Theta^*) \|^2 \\
& \leq & \gamma_1 \gamma_2 \| (\Theta^i - \Theta^*) \|^2
\end{IEEEeqnarray*}
where $\gamma_{1,i}$ is as in Corollary \ref{corcose}, $\gamma_1 = \max_i \{ \gamma_{1,i} \}$ and $\gamma_2= \| \bar \Theta_u^* \|$. Notice that, since $\Theta^i$ converges by Theorem \ref{thcvg}, $\gamma_1$ is well defined.
\end{proof}

We have shown that Algorithm 1 successfully solves the LQR control problem using only measured data from the system. To execute Algorithm 1, knowledge of an initial stabilizing matrix $K^0$ is required. In the following section we show how to compute a deadbeat controller in a data-based fashion.

\section{A Data-Based Initial Stabilizing Controller}
\label{secdb}

We are interested in determining an initial stabilizing matrix $K^0$, without knowledge of the system model, and without losing the important efficiency advantages of Algorithm 1 that we analyze later in Section \ref{seccomp}. In \cite[Theorem~21]{Waardeetal2:20}, it is shown that a deadbeat controller can be obtained for an unknown system using pole-placement strategies on a pair of matrices $(\bar A, \bar B)$ constructed from data. Note that this method does not imply that the system matrices $(A,B)$ are identified. Since solving this pole-placement problem is not trivial due to the nature of the data matrices $\bar A$ and $\bar B$, and since no further discussion is provided in \cite{Waardeetal2:20}, we show for the convenience of the reader a series of steps that yields the corresponding result. Consider Algorithm 2 below.

The algorithm starts by defining the matrices $h_0(x)$, $h_1(x)$ and $h_0(u)$ in (\ref{h0})-(\ref{hu}) using persistently excited data. This information is then used to define the matrices $\bar A$ and $\bar B$ of a fictitious system. As these matrices are known, "model-based" strategies can be employed to design a feedback gain $\bar H$ to place all the poles of $\bar A - \bar B \bar H$ at zero. In \cite{Waardeetal2:20} it is proven that $\bar H$ can be used to determine the actual deadbeat gain $K_{db}$ for the true system (\ref{linsys}), as stated in the following theorem.

\begin{figure}[h]
\hrule
{\bf Algorithm 2: Data-based Deadbeat Control}
{\hrule \small
	\begin{algorithmic}[1]
		\Procedure{}{}
		\State Collect $N \geq (m+1)(n+1)-1$ samples of data $\{ x_k, u_k \}_{k=0}^{N-1}$ by applying a PE input of order $n+1$ to the system.
		\State Define the matrices
		\begin{equation}
			h_0(x)=\left[ \begin{array}{cccc} x_0 & x_1 & \cdots & x_{N-2} \end{array} \right] \in \mathbb{R}^{n \times (N-1)},
			\label{h0}
		\end{equation}
		\begin{equation}
			h_1(x)=\left[ \begin{array}{cccc} x_1 & x_2 & \cdots & x_{N-1} \end{array} \right] \in \mathbb{R}^{n \times (N-1)},
			\label{h1}
		\end{equation}
		\begin{equation}
			h_0(u)=\left[ \begin{array}{cccc} u_0 & u_1 & \cdots & u_{N-2} \end{array} \right] \in \mathbb{R}^{m \times (N-1)}.
			\label{hu}
		\end{equation}
		\State Let $F$ be the Moore-Penrose pseudoinverse of $h_0(x)$, and let $G$ be a basis for the nullspace of $h_0(x)$. Define the matrices $\bar A = h_1(x)F$ and $\bar B = h_1(x)G$.
		\State Define $\bar B_F$ as a full-column-rank matrix constructed from a set of linearly independent columns of $\bar B$, such that
		\begin{equation*}
			\text{rank}(\bar B_F) = \text{rank}(\bar B).
		\end{equation*}
		\State Transform the system $(\bar A, \bar B_F)$ into a MIMO controller canonical form, $(A_C, B_C)$, using a transformation matrix $T$.
		\State Design a deadbeat controller, $H_C$, for the system $(A_C, B_C)$, i.e., let $A_C-B_CH_C$ be nilpotent.
		\State Compute the matrix $H_F= H_C T$, with $T$ as in Step 6.
		\State Construct the matrix $\bar H$ by adding zero rows to $H_F$, in the same row positions as the position of the columns of $\bar B$ that were removed to form $\bar B_F$, such that
		\begin{equation}
			\bar B_F H_F = \bar B \bar H.
			\label{bhz}
		\end{equation}
		\State Compute the deadbeat gain $K_{db}= - h_0(u)(F-G \bar H)$.
		\EndProcedure
		\hrule
	\end{algorithmic}
}
\end{figure}

\begin{thm}[\cite{Waardeetal2:20}]
The matrix $K_{db}$ that results from Algorithm~2 is a deadbeat control gain for the system (\ref{linsys}), i.e., all the eigenvalues of $A-BK_{db}$ are equal to zero.
\end{thm}

The main obstacle in the application of this procedure is the fact that the matrix $\bar B$ does not have full column rank. This inconvenience can be addressed by neglecting the linearly dependent columns of $\bar B$, and constructing a full-column-rank matrix $\bar B_F$ using the remaining columns. Later, when we have determined a feedback matrix $\bar H_F$ for the system $(\bar A, \bar B_F)$, we only need to add zero-rows to $\bar H_F$ in the appropriate positions to make (\ref{bhz}) hold. The full-rank matrix $\bar B_F$ can be used to solve the pole-placement problem for MIMO systems. We propose the method of transforming the system into a controllable form as described in \cite{Luen:67}. Using this controllable form, designing the desired feedback gain $\bar H_F$ is straightforward.

Since the deadbeat controller is stable, it can be used as the initial stabilizing policy for Algorithm 1. From this starting point, Algorithm 1 computes the optimal contol gain $K^*$ that minimizes the desired cost function (\ref{cost}).

\section{Robustness Properties of the Q-Learning Algorithm}
\label{secrob}

Besides the initial stabilizing controller, another practical consideration for the implementation of Algorithm 1 is the presence of noise in the state data $x_k$ obtained from the system. In this section, we show that Algorithm 1 presents inherent robustness properties against noise in the measurements. 

Assume that the system states evolve according to the dynamics (\ref{linsys}), but that the available measurements of the states are corrupted by noise, such that 
\begin{equation}
	\chi_k = x_k + w_k,
	\label{noissig}
\end{equation}
where $w_k \in \mathbb{R}^n$ is an unknown, bounded noise term. We now define the noise-corrupted vectors $\hat z_k = [\chi_k^\top \,\,\, u_k^\top ]^\top$ and $\hat \zeta_{i,k+1} = [\chi_{k+1}^\top \,\,\, -(K^i \chi_{k+1})^\top ]^\top$.

The first question that arises is whether Lemma \ref{lemwil} still holds when a PE input is applied to the system and noisy data is collected from it. That is, whether the rank condition
\begin{equation}
	\text{rank} \left( \left[ \begin{array}{c} H_1(\chi_{[0,N-1]}) \\ H_1(u_{[0,N-1]}) \end{array} \right] \right) = \eta
	\label{wlranknois}
\end{equation}
holds, where $\eta = n+m$ (compare (\ref{rankz})). The following lemma shows that this is the case if the magnitude of the noise terms $w_k$ is small enough.

\begin{lem}
	\label{lemnoispe}
	Let the data $\{ \chi_k \}_{k=0}^{N-1}$, with $\chi_k$ as in (\ref{noissig}), be collected from a controllable system (\ref{linsys}) when an input $\{ u_k \}_{k=0}^{N-1}$ which is PE of order $n+1$ is applied to it. If the noise signal $\{ w_k \}_{k=0}^{N-1}$ is such that
	\begin{equation}
		\| H_1(w_{[0,N-1]}) \|_2 < \sigma_{\min} \left( \left[ \begin{array}{c} H_1(x_{[0,N-1]}) \\ H_1(u_{[0,N-1]}) \end{array} \right] \right)
		\label{noisrankcond}
	\end{equation}
	then (\ref{wlranknois}) holds.
\end{lem}
\begin{proof}
	Consider a full rank matrix $M$. A necessary condition for the matrix $M + \Delta$ to be rank deficient is known to be $\| \Delta \| \geq \sigma_{\min}(M)$, which holds for the Frobenius norm \cite{GoLo:80} and for the 2-norm \cite{DaDaVe:11}. The proof is completed by writing
	\begin{equation*}
		\left[ \begin{array}{c} H_1(\chi_{[0,N-1]}) \\ H_1(u_{[0,N-1]}) \end{array} \right] = \left[ \begin{array}{c} H_1(x_{[0,N-1]}) \\ H_1(u_{[0,N-1]}) \end{array} \right] + \left[ \begin{array}{c} H_1(w_{[0,N-1]}) \\ 0 \end{array} \right].
	\end{equation*}
\end{proof}

Now, Algorithm 1 is executed using the data $\{ \chi_k, u_k \}_{k=0}^{N-1}$. Since we solve the Bellman equation (\ref{qthi}) using the generalized Sylvester equation (\ref{thmatzy}), define the noise-corrupted matrices (compare with (\ref{matrixz})-(\ref{matrixy}))
\begin{equation}
	\hat Z := \left[ \begin{array}{cccc} \hat z_{k_1} & \hat z_{k_2} & \cdots & \hat z_{k_\eta} \end{array} \right]
	\label{matrixzw}
\end{equation}
and 
\begin{equation}
	\hat Y_i := \left[ \begin{array}{cccc} \hat \zeta_{i,k_1+1} & \hat \zeta_{i,k_2+1} & \cdots & \hat \zeta_{i,k_\eta+1} \end{array} \right].
	\label{matrixyw}
\end{equation}
Moreover, from the noise signal $\{ w_k \}_{k=0}^{N-1}$ it is useful to define
\begin{equation*}
	W_- := \left[ \begin{array}{cccc} w_{k_1} & w_{k_2} & \cdots & w_{k_\eta} \end{array} \right]
\end{equation*}
and
\begin{equation*}
	W_+ := \left[ \begin{array}{cccc} w_{k_1+1} & w_{k_2+1} & \cdots & w_{k_\eta+1} \end{array} \right].
\end{equation*}

Using (\ref{matrixzw})-(\ref{matrixyw}), we replace (\ref{thmatzy}) by
\begin{equation}
\hat Z^\top \hat \Theta^{i+1} \hat Z = \hat Z^\top \bar Q \hat Z +  \hat Y_i^\top \hat \Theta^{i+1} \hat Y_i.
\label{thmatzyw}
\end{equation}
Later, in Theorem \ref{thstaw}, we provide conditions on the noise signal that guarantee the existence of a unique solution $\hat \Theta^{i+1}$ of (\ref{thmatzyw}). The solution $\hat \Theta^{i+1}$ of (\ref{thmatzyw}) differs from the solution $\Theta^{i+1}$ of (\ref{thmatzy}) due to the effect of the noise.

Notice that, unlike in the noiseless case, $\hat \zeta_{i,k+1} \neq \Phi_i \hat z_k$. However, notice that 
\begin{IEEEeqnarray*}{rCl}
\hat \zeta_{i,k+1} & = & \left[ \begin{array}{c} x_{k+1}+w_{k+1} \\ -K^i(x_{k+1}+w_{k+1}) \end{array} \right] + \Phi_i \left[ \begin{array}{c} w_{k} \\ 0 \end{array} \right] - \Phi_i \left[ \begin{array}{c} w_{k} \\ 0 \end{array} \right] \\
& = & \Phi_i \hat z_k + \bar K_i \bar w_k,
\end{IEEEeqnarray*}
with $\bar K_i = [ I \quad -(K^{i})^\top ]^\top$, and
\begin{equation}
\bar w_k = w_{k+1} - A w_k.
\label{wbar}
\end{equation}
This expression implies
\begin{equation}
	\hat Y_i = \Phi_i \hat Z + \bar K_i \bar W
	\label{ywexp}
\end{equation}
with $\bar W = W_+ - A W_-$. This allows us to write (\ref{thmatzyw}) as
\begin{equation}
		\hat Z^\top \hat \Theta^{i+1} \hat Z = \hat Z^\top \bar Q \hat Z +  \hat Z \Phi_i^\top \hat \Theta^{i+1} \Phi_i \hat Z + \mathcal{E}_i
		\label{noiseq}
\end{equation}
where
\begin{multline}
		\mathcal{E}_i = \hat Z^\top S^\top \bar K_i^\top \hat \Theta^{i+1} \bar K_i \bar W + \bar W^\top \bar K_i^\top \hat \Theta^{i+1} \bar K_i S \hat Z \\
		+ \bar W^\top \bar K_i^\top \hat \Theta^{i+1} \bar K_i \bar W
	\label{epsw}
\end{multline}
and $S=[A \quad B]$. Here we have used the fact that $\Phi_i= \bar K_i S$.

For a given matrix $K^i$, we can use (\ref{noiseq}) to quantify the difference between $\hat \Theta^{i+1}$ in (\ref{thmatzyw}) and $\Theta^{i+1}$ in (\ref{thmatzy}),  in terms of $\bar W$ as follows. Define $\Delta \Theta^{i+1} = \hat \Theta^{i+1} - \Theta^{i+1}$. Now, rewrite (\ref{thmatzy}) as 
\begin{multline}
		Z^\top \hat \Theta^{i+1} Z = Z^\top \bar Q Z +  Z^\top \Phi_i^\top \hat \Theta^{i+1} \Phi_i Z \\
		+ Z^\top \Delta \Theta^{i+1} Z - Z^\top \Phi_i^\top \Delta \Theta^{i+1} \Phi_i Z.
		\label{tempeq}
\end{multline}
Since both $Z$ and $\hat Z$ are nonsingular matrices (by assuming that (\ref{noisrankcond}) holds), then by multiplying (\ref{tempeq}) from the left by $\hat Z^\top (Z^{-1})^\top$ and from the right by $Z^{-1} \hat Z$, we obtain the equivalent equation
\begin{multline*}
	\hat Z^\top \hat \Theta^{i+1} \hat Z = \hat Z^\top \bar Q \hat Z +  \hat Z^\top \Phi_i^\top \hat \Theta^{i+1} \Phi_i \hat Z \\
	+ \hat Z^\top \Delta \Theta^{i+1} \hat Z - \hat Z^\top \Phi_i^\top \Delta \Theta^{i+1} \Phi_i \hat Z.
\end{multline*}
Comparing this equation with (\ref{noiseq}), we get 
\begin{equation}
		\hat Z^\top \Delta \Theta^{i+1} \hat Z - \hat Z^\top \Phi_i^\top \Delta \Theta^{i+1} \Phi_i \hat Z = \mathcal{E}_i
		\label{dthw}
\end{equation}
relating $\Delta \Theta^{i+1}$ with $\bar W$ via $\mathcal{E}_i$ in (\ref{epsw}). The following lemma provides conditions for the existence of a unique solution $\hat \Theta^{i+1}$ to (\ref{thmatzyw}), as well as for $K^{i+1}$ to be stabilizing at each iteration of Algorithm 1, in terms of $\Delta \Theta^{i+1}$. Subsequently, in Theorem~\ref{thstaw}, we use (\ref{dthw}) to express this condition in terms of the noise matrix $\bar W$.

\begin{lem}
Let the condition (\ref{noisrankcond}) hold, and let the matrix $K^i$ stabilize the system (\ref{linsys}). If $\Delta \Theta^{i+1}$ in (\ref{tempeq}) is such that
\begin{equation}
		\| \Phi_i^\top \Delta \Theta^{i+1}  \Phi_i -\Delta \Theta^{i+1} \|_2 < \lambda_{\min}(\bar Q),
		\label{constab1}
\end{equation}
then (\ref{thmatzyw}) has a unique solution $\hat \Theta^{i+1}$, and the matrix $K^{i+1}=(\hat \Theta_{uu}^{i+1})^{-1}\hat \Theta_{ux}^{i+1}$ also stabilizes the system (\ref{linsys}).
\label{lemstaw}
\end{lem}
\begin{proof}
First, notice that we can express (\ref{thmatzyw}) as in (\ref{noiseq}). Now, define $\bar Q_2 := \bar Q + (\hat Z^{-1})^\top \mathcal{E}_i \hat Z^{-1}$, where $\hat Z^{-1}$ exists by Lemma~\ref{lemnoispe}. This allows us to write (\ref{noiseq}) as 
\begin{equation*}
	\hat Z^\top \hat \Theta^{i+1} \hat Z = \hat Z^\top \bar Q_2 \hat Z +  \hat Z^\top \Phi_i^\top \hat \Theta^{i+1} \Phi_i \hat Z.
\end{equation*}
Since $\hat Z$ is nonsingular, this is equivalent to 
\begin{equation*}
	\hat \Theta^{i+1} = \bar Q_2 +  \Phi_i^\top \hat \Theta^{i+1} \Phi_i,
\end{equation*}
which, by stability of $K^i$, has a unique solution $\hat \Theta^{i+1}$ if $\bar Q_2 \succ 0$ (compare Corollary \ref{corsol}). Moreover, following a similar procedure as in Theorem \ref{thstab}, this equation also shows the stabilizing property of $K^{i+1}$, as long as $\bar Q_2 \succ 0$. To show that $\bar Q_2 \succ 0$ holds, rewrite (\ref{dthw}) as  
\begin{equation}
	\Delta \Theta^{i+1} - \Phi_i^\top \Delta \Theta^{i+1}  \Phi_i = (\hat Z^{-1})^\top \mathcal{E}_i \hat Z^{-1}
	\label{deleps}
\end{equation}
and, thus, $\bar Q_2 = \bar Q + \Delta \Theta^{i+1} - \Phi_i^\top \Delta \Theta^{i+1}  \Phi_i$. Finally, the condition (\ref{constab1}) guarantees the positive definiteness of $\bar Q_2$.
\end{proof}

\begin{thm}
Let the condition (\ref{noisrankcond}) hold and let the matrix $K^i$ stabilize the system (\ref{linsys}). If $\bar W$ in (\ref{noiseq})-(\ref{epsw}) is such that
\begin{multline}
		\| \hat \Theta^{i+1} \|_2 \| \bar K_i \|_2^2 \| \hat Z^{-1} \|_2 \| \bar W \|_2 \\
		\times \left( 2 \| S \|_2 + \| \hat Z^{-1} \|_2 \| \bar W \|_2 \right) < \lambda_{\min}(\bar Q),
\label{constab2}
\end{multline}
then (\ref{thmatzyw}) has a unique solution $\hat \Theta^{i+1}$, and the matrix $K^{i+1}=(\hat \Theta_{uu}^{i+1})^{-1}\hat \Theta_{ux}^{i+1}$ also stabilizes the system (\ref{linsys}).
\label{thstaw}
\end{thm}
\begin{proof}
From (\ref{deleps}), we have $\| \Delta \Theta^{i+1} - \Phi_i^\top \Delta \Theta^{i+1} \Phi_i \|_2 = \| (\hat Z^{-1})^\top \mathcal{E}_i \hat Z^{-1} \|_2$. Noticing that the left-hand side of (\ref{constab2}) is an upper bound for $\| (\hat Z^{-1})^\top \mathcal{E}_i \hat Z^{-1} \|_2$ (see (\ref{epsw})), the proof is completed by Lemma~\ref{lemstaw}.
\end{proof}

In the following theorem, we show that the condition (\ref{constab2}) also guarantees the convergence of Algorithm 1 in the presence of noise, in the sense that $\hat \Theta^i \succeq \hat \Theta^{i+1}$.

\begin{thm}
	Let the conditions (\ref{noisrankcond}) and (\ref{constab2}) hold, and let the matrix $K^i$ stabilize the system (\ref{linsys}). Then, $\hat \Theta^i \succeq \hat \Theta^{i+1}$ for all iterations of Algorithm~1.
	\label{thmconvw}
\end{thm}
\begin{proof}
	Using (\ref{ywexp}) and the fact that $\hat Z$ is nonsingular, (\ref{thmatzyw}) can be written as
	\begin{equation*}
		\hat \Theta^{i+1} = \bar Q +  (\Phi_i + \bar K_i \bar W \hat Z^{-1})^\top \hat \Theta^{i+1} (\Phi_i + \bar K_i \bar W \hat Z^{-1}).
	\end{equation*}
Defining $\hat \Phi_i = \Phi_i + \bar K_i \bar W \hat Z^{-1}$, this expression becomes
\begin{equation}
	\hat \Theta^{i+1} = \bar Q +  \hat \Phi_i^\top \hat \Theta^{i+1} \hat\Phi_i.
	\label{hatphi}
\end{equation}
Since $\Phi_i = \bar K_i [A \quad B]$, then $\hat \Phi_i = \bar K_i ([A \quad B] + \bar W \hat Z^{-1}) = \bar K_i [\hat A \quad \hat B]$, where $[\hat A \quad \hat B] := [A \quad B] + \bar W \hat Z^{-1}$. Finally, since (\ref{hatphi}) has a unique solution $\hat \Theta^{i+1}$ by Theorem \ref{thstaw}, it can be used to follow the steps in Lemma \ref{lemide2}, Corollary \ref{coride3} and Lemma \ref{lemconv1} (replacing the matrices $A$ and $B$ by $\hat A$ and $\hat B$), to obtain the desired result. 
\end{proof}

\begin{rem}
	Notice that the matrices $\hat A$ and $\hat B$ defined in the proof of Theorem \ref{thmconvw} are used only for analysis purposes and do not correspond to a model identification step in Algorithm~1. The algorithm only requires the solution of the data-based equation (\ref{thmatzyw}).
\end{rem}

\begin{rem}
We highlight the fact that the condition (\ref{constab2}) depends on $\bar W$ being small enough. From (\ref{wbar}), this implies that noise samples $w_{k+1}$ of large magnitude would not affect the performance of Algorithm 1 if $w_{k+1}=Aw_k$. A similar effect was observed in \cite{DeTi:20}.
\end{rem}

In many practical applications, it may be of interest to express (\ref{constab2}) as a condition on the magnitude of the samples $w_k$, instead of $\bar w_k$. This is possible by expressing $\| \bar W \|_2 \leq \| W_+ \|_2 + \| A \|_2 \| W_- \|_2$. Assume now that the noise samples are bounded such that $\| W_- \|_2, \, \| W_+ \|_2 \leq W_{\max}$. Then, we find the bound $\| \bar W \|_2 \leq (1 + \| A \|_2) W_{\max}$. Finally, the condition (\ref{constab2}) is replaced by
\begin{multline*}
	\| \hat \Theta^{i+1} \|_2 \| \bar K_i \|_2^2 \| \hat Z^{-1} \|_2 (1 + \| A \|_2) W_{\max} \\
	\times \left( 2 \| S \|_2 + \| \hat Z^{-1} \|_2 (1 + \| A \|_2) W_{\max} \right) < \lambda_{\min}(\bar Q),
\end{multline*}

In order to be able to compute the bound (\ref{constab2}), an upper bound on $\| S \|_2 = \| [A \quad B] \|_2$ is assumed to be known. Moreover, since $\hat \Theta^i \succeq \hat \Theta^{i+1}$ by Theorem \ref{thmconvw}, we can determine a bound on $\hat \Theta^i$ for all $i$ after running one iteration of the algorithm. Finally, a bound on $\| \bar K_i \|_2$ needs to be known. This issue is addressed in the following.

In Lemma \ref{lemstaw}, we have used the condition $\bar Q + \Delta \Theta^{i+1} - \Phi_i^\top \Delta \Theta^{i+1} \Phi_i \succ 0$ to guarantee stability. If, instead, we make the more conservative requirement 
\begin{equation}
	\bar Q + \Delta \Theta^{i+1} - \Phi_i^\top \Delta \Theta^{i+1} \Phi_i \succ I,
	\label{newcond}
\end{equation}
we can determine a bound as (\ref{constab2}) that does not depend on $\| \bar K_i \|$. Notice that, in this case, the user defined matrix $\bar Q$ would be restricted to be $\bar Q \succ I$. To determine this bound, consider (\ref{tempeq}) and use the condition (\ref{newcond}) to get
\begin{equation*}
	Z^\top \hat \Theta^{i+1} Z \succ Z^\top Z +  Z^\top \Phi_i^\top \hat \Theta^{i+1} \Phi_i Z \succeq Z^\top Z
\end{equation*}
and, therefore, $\hat \Theta^{i+1} -I \succ 0$. For notation convenience, in the following we shift the index $i+1$ for $i$. Partitioning the matrix $\hat \Theta^{i}$ as 
\begin{equation*}
	 \hat \Theta^{i}= \left[ \begin{array}{cc} \hat \Theta_{xx}^i & (\hat \Theta_{xu}^i)^\top \\ \hat \Theta_{xu}^i & \hat \Theta_{uu}^i \end{array} \right],
\end{equation*}
we notice that this directly implies $\hat \Theta_{uu}^i -I \succ 0$. Moreover, note that $\hat \Theta^{i}$ can be expressed in terms of its Schur complement as
\begin{multline}
	\hat \Theta^{i} = \left[ \begin{array}{cc} I & K^{i \top} \\ 0 & I \end{array} \right] \left[ \begin{array}{cc} \hat \Theta_{xx}^i - (\hat \Theta_{xu}^i)^\top (\hat \Theta_{uu}^i)^{-1} \hat \Theta_{xu}^i & 0 \\ 0 & \hat \Theta_{uu}^i \end{array} \right]  \\
	\times \left[ \begin{array}{cc} I & 0 \\ K^i  & I \end{array} \right],
	\label{thschur}
\end{multline}
where $K^i = (\hat \Theta_{uu}^i)^{-1} \hat \Theta_{xu}^i$. From (\ref{thschur}) it is easy to show that $\hat \Theta^i -I \succ 0$ implies $\hat \Theta_{xx}^i - (\hat \Theta_{xu}^i)^\top (\hat \Theta_{uu}^i)^{-1} \hat \Theta_{xu}^i -I \succ 0$.

Therefore, $\hat \Theta^i -I \succ 0$ implies
\begin{IEEEeqnarray*}{rCl}
	\hat \Theta^{i} & = & \left[ \begin{array}{cc} I & K^{i \top} \\ 0 & I \end{array} \right] \left[ \begin{array}{cc} \hat \Theta_{xx}^i - (\hat \Theta_{xu}^i)^\top (\hat \Theta_{uu}^i)^{-1} \hat \Theta_{xu}^i & 0 \\ 0 & \hat \Theta_{uu}^i \end{array} \right] \\
	& & \IEEEeqnarraymulticol{1}{r}{\times \left[ \begin{array}{cc} I & 0 \\ K^i  & I \end{array} \right]} \\
	& \succ & \left[ \begin{array}{cc} I & K^{i \top} \\ 0 & I \end{array} \right] \left[ \begin{array}{cc} I & 0 \\ K^i  & I \end{array} \right].
\end{IEEEeqnarray*}
Finally, we obtain
\begin{equation*}
	\| \hat \Theta^{i} \|_2 > \left\| \left[ \begin{array}{cc} I & K^{i \top} \\ 0 & I \end{array} \right] \left[ \begin{array}{cc} I & 0 \\ K^i  & I \end{array} \right] \right\|_2  >  \| \bar K_i \|_2^2. 
\end{equation*}
Therefore, we can substitute $\| \bar K_i \|_2^2$ by $\| \hat \Theta^{i} \|_2$ and $\bar Q$ by $\bar Q -I$ in (\ref{constab2}). Since, by Theorem \ref{thmconvw}, $\hat \Theta^i \succeq \hat \Theta^{i+1}$, then $\| \hat \Theta^1 \|_2$ is a uniform bound for all iterations $i$. Condition (\ref{constab2}) is hence replaced by
\begin{equation}
	\| \hat \Theta^{1} \|_2^2 \| \hat Z^{-1} \|_2 \| \bar W \|_2 \left( 2 \| S \|_2 + \| \hat Z^{-1} \|_2 \| \bar W \|_2 \right) < \lambda_{\min}(\bar Q).
	\label{constab3}
\end{equation}
We emphasize the intuition behind the conditions (\ref{constab2}) and (\ref{constab3}). These expressions provide the qualitative result that they are always satisfied for small enough values of $\| \bar W \|_2$ and, therefore, the algorithm preserves its convergence and stability properties for small magnitudes of the noise. Moreover, arbitrarily small magnitudes of the noise imply convergence to a control gain arbitrarily close to the optimal controller, as stated in the following remark.

\begin{rem}
In the presence of noise, the conditions (\ref{noisrankcond}) and (\ref{constab3}) allow Algorithm 1 to converge to a solution $\hat \Theta^*$ that is related to the true LQR value $\Theta^*$ by $\hat \Theta^* - \Theta^* = \Delta \Theta^*$, where $\Delta \Theta^*$ satisfies (see (\ref{dthw}))
\begin{equation*}
	 \Delta \Theta^{*} - \Phi_*^\top \Delta \Theta^{*} \Phi_* = (\hat Z^\top)^{-1} \mathcal{E}^* \hat Z^{-1}
\end{equation*}
where $\mathcal{E}^*$ is as in (\ref{epsw}) with $\hat \Theta^{i+1}$ replaced by $\hat \Theta^*$, and similarly for $\bar K_i$. Moreover, from (\ref{hatphi}), it can also be noted that the final solution $\hat \Theta^*$ corresponds to the LQR solution for a system with matrices $(\hat A, \hat B)$, where $[\hat A \quad \hat B] := [A \quad B] + \bar W \hat Z^{-1}$. Notice that $\hat \Theta^* \rightarrow \Theta^*$ as $w_k \rightarrow 0$.
\end{rem}

Currently, Algorithm 1 is one of the increasing number of data-based methods to solve this problem. In the next section, we show that the low computational complexity associated with Algorithm 1 makes it an attractive alternative in this area.

\section{Efficiency Comparison with Existing Data-Based Algorithms}
\label{seccomp}

In this section, we compare the computational complexity of Algorithm 1 with that of existing model-free algorithms in the literature that address the discrete-time LQR problem. In particular, we compare Algorithm 1 to other Q-learning algorithms and to data-driven methods based on LMIs. We remark that this computational efficiency analysis is performed in terms of the number of operations required to complete the algorithm, and not in the sense of the amount of data required to be collected from the system. We consider such a sample complexity analysis of these algorithms to be an interesting area for future research. Here, we first show that the off-policy iterative algorithm in \cite{KiLeJi:17} requires the solution of a larger set of equations than Algorithm 1. Then, we find that completing Algorithm 1 requires a lower order of the number of operations than the efficient LMI method presented in \cite{PeTe:21}.

\subsection{An existing off-policy algorithm}
In \cite{KiLeJi:17}, a model-free off-policy algorithm (\hspace{1sp}\cite[Algorithm~3]{KiLeJi:17}) is designed. It can be noted that \cite{KiLeJi:17} studies the $H_\infty$ control problem and, therefore, some discussion is required before comparing it to our Algorithm~1. The problem with solving the $H_\infty$ problem in a data-based fashion is that measurements of the noise signal need to be considered as part of the collected data. Of course, the noise is unknown at every instant, making the application of this algorithm difficult in practice. However, \cite[Algorithm~3]{KiLeJi:17} can successfully solve the LQR problem if the system is not affected by disturbances.

The method used in \cite{KiLeJi:17} to design their algorithm consists in expressing the Hamilton-Jacobi equation (rather than the Q-function) as a set of linear equations. Moreover, the persistence of excitation condition proposed there is not easily verifiable, differing from our Willems'-lemma-based discussion. In \cite{KiLeJi:17}, it is explicitly stated that a linear set of equations with $n^2+m^2+m_d^2+2mm_d+n(m+m_d)$ unknowns must be solved for the $H_\infty$ problem, where $m_d$ is the number of entries in the disturbance vector. The corresponding LQR version of their algorithm therefore has $n^2+m^2+nm$ unknowns. In contrast, Algorithm 1 in this paper requires the solution of the set of equations (\ref{qthi}) with $\frac{1}{2}(n+m)(n+m+1)$ unknowns. This reduced amount of parameters guarantees a decreased complexity that becomes more significant as the numbers of states and inputs of the system increase. In Section \ref{secsimul}, the significant difference between the performance of these algorithms is evidenced using simulation examples.

\subsection{Existing LMI methods}
In recent years, many data-based LQR controllers have been proposed that use LMIs in their formulation \cite{WaMe:21,PeTe:21,Berbeetal:20,WaCaMe:21}. Optimization problems with LMIs are solved using \emph{interior-point} methods, that are known to solve such optimization problems in $\mathcal{O}(\rho^{2.75}L^{1.5})$ arithmetic operations \cite{Boydetal:94}, where $\rho$ is the number of unknowns and $L$ is the number of LMIs that constrain the problem. In \cite{PeTe:21}, a low complexity LMI method for the LQR problem is proposed, with a total of $nN+\frac{m(m+1)}{2}$ unknowns, where $N \geq (n+1)m+n$ is the amount of data collected. This implies an complexity of $\mathcal{O}((n^{2.75}N^{2.75} + m^{5.5})L^{1.5})$ operations, where $L=4$.

In contrast, the complexity of the Q-learning algorithm proposed in this paper is determined by three sources: the selection in Step 3 of $\eta$ linearly independent vectors from the collected data, the solution of the set of equations (\ref{qthi}) (which is equivalent to solving (\ref{thmatzy}), see Theorem \ref{thcons}) at every iteration, and the number of iterations until convergence. First, a simple method for selecting the linearly independent vectors is to organize the data collected in a matrix as the one shown in (\ref{rankz}), and transform this matrix into row echelon form. This procedure is performed only once in Algorithm~1 and uses Gaussian elimination which, for a $(n+m) \times N$ matrix, has a complexity of $\mathcal{O}(N(n+m)^2)$ operations \cite{Strang:06}, with $N \geq (n+1)m+n$. Next, algorithms to solve equations of the form (\ref{thmatzy}) require $\mathcal{O}((n+m)^3)$ operations (see, e.g., \cite{SaCha:20}). 

Finally, Theorem \ref{thqconv} shows that the algorithm has quadratic order of convergence. Notice from (\ref{quconv}) that the final rate of convergence depends on the factor $\gamma$ which, as shown in the proof of Theorem \ref{thqconv}, depends on the system matrices $A$ and $B$, as well as on the iteration matrices $K^i$.  However, the order of the error at every iteration decreases quadratically, independently of these factors. This implies that the order of the number of iterations required to reach a tolerance error does not increase as a direct effect of $n$ and $m$. Since the number of iterations is a constant factor of certain order of magnitude, and since $N > n+m$, the order of the number of operations required to complete Algorithm 1 is $\mathcal{O}(N(n+m)^2)$. Notice that the computational advantage of Algorithm 1 becomes more significant as the dimension of the system and the amount $N$ of data collected increase. This theoretical analysis is also tested by means of numerical simulations in the following section.

\section{Simulation results}
\label{secsimul}

In this section, we test the theoretical results presented in Section \ref{secrob} and Section \ref{seccomp}. First, we perform efficiency comparisons between Algorithm 1 and two other data-based algorithms that solve the LQR problem.

\subsection{Performance comparison with data-based methods}
In Section \ref{seccomp}, we showed that Algorithm 1 is more efficient than other model-free reinforcement learning algorithms, as well as algorithms that require the solution of LMIs. In this subsection, we test in simulation these theoretical results. Here, Algorithm 1 is compared with the LMI method in \cite{PeTe:21} and with the alternative RL method in \cite{KiLeJi:17}. For brevity, here we refer to Algorithm 1 as QL, the method in \cite{PeTe:21} as LMI, and the algorithm in \cite{KiLeJi:17} as RL. Many different simulations are performed for linear systems of the form (\ref{linsys}), where the matrices $A$ and $B$ are generated randomly with entries ranging from $-1$ to $1$, and the matrix $A$ is not necessarily stable. To test the computational complexity at different dimensions of the system, our simulations consider dimensions from $n=3$ to $n=50$. A total of 100 simulations are executed for each value of $n$. In all cases, we keep the fixed value $m=2$.

For each of these linear systems, an appropriate PE input $u_k$ is designed and applied to the system for data collection. Since we aim to test computational complexity, we let each algorithm use the minimum amount of data that they require to run. QL is run for exactly 10 iterations, regardless of the system dimension. To solve the generalized Sylvester equation (\ref{thmatzy}), we use the command \textit{dlyap} in Matlab. Regarding the LMI method, we use CVX from Matlab to solve the optimization problem. Finally, due to its slower convergence, the algorithm RL is run for 15 iterations, regardless of the system dimension.  When the matrix $A$ is unstable, Algorithm 2 in Section \ref{secdb} is used to determine an initial stabilizing policy for QL and RL. For all simulations, Matlab R2020b was used on an Intel i7-10875H (2.30 GHz) with 16 GB of memory.

Table \ref{tab1} compares the performances of the three algorithms after 100 simulations (for each value of $n$) in terms of the average error norm $\varepsilon = \| K^* - K^c \|$, where $K^*$ is the actual optimal solution of the problem in hand and $K^c$ is the matrix gain obtained upon convergence of the algorithm. Table \ref{tab2} shows the comparison between the average time to completion (in seconds) for each of the tested methods. It is observed that the proposed QL algorithm outperforms the other two methods with increasing efficiency as the dimension of the system grows.

\begin{table}
\centering
\begin{tabular}{|c|c c c|} 
	\hline
	\multirow{2}{1.2cm}{\centering Dimension $n$} & \multicolumn{3}{c|}{Average error $\varepsilon$} \\
	& QL & LMI & RL \\[0.5ex]
	\hline
	$3$ & $9.744 \times 10^{-15}$ & $7.003 \times 10^{-8}$ & $1.504 \times 10^{-7}$ \\ 
	$5$ & $7.767 \times 10^{-13}$ & $1.087 \times 10^{-7}$ & $1.312 \times 10^{-7}$ \\
	$10$ & $7.447 \times 10^{-12}$ & $1.548 \times 10^{-6}$ & $7.126 \times 10^{-6}$ \\
	$20$ & $6.595 \times 10^{-11}$ & $2.396 \times 10^{-5}$ & $1.073 \times 10^{-5}$ \\
	$50$ & $1.902 \times 10^{-10}$ & $0.760$ & $6.615 \times 10^{-4}$ \\ [1ex] 
	\hline
\end{tabular}
\vspace{4pt}
\caption{Error comparison between Algorithm 1 (QL) and the algorithms in \cite{PeTe:21} (LMI) and \cite{KiLeJi:17} (RL).}
\label{tab1}
\end{table}
\begin{table}
	\centering
	\begin{tabular}{|c|c c c|} 
		\hline
		\multirow{2}{1.2cm}{\centering Dimension $n$} & \multicolumn{3}{c|}{Average run-time $(s)$} \\
		& QL & LMI & RL \\[0.5ex]
		\hline
		$3$ & $8.639 \times 10^{-4}$ & $0.293$ & $0.009$ \\ 
		$5$ & $9.976 \times 10^{-4}$ & $0.378$ & $0.015$ \\
		$10$ & $0.001$ & $0.599$ & $0.045$ \\
		$20$ & $0.003$ & $4.963$ & $0.236$ \\
		$50$ & $0.015$ & $84.345$ & $40.05$ \\ [1ex] 
		\hline
	\end{tabular}
\vspace{4pt}
	\caption{Run-time comparison between Algorithm 1 (QL) and the algorithms in \cite{PeTe:21} (LMI) and \cite{KiLeJi:17} (RL).}
	\label{tab2}
\end{table}

\subsection{Performance comparison with model-based methods and noisy measurements}
Recent publications (\hspace{1sp}\cite{DoCoMa:22,KrPa:21}) have made a thorough analysis comparing different data-based control approaches. One possibility is to design stabilizing controllers directly from measured data (direct data-based control). Alternatively, the measured data can be used first to identify the model of the system, and then model-based design is used to determine a controller (indirect data-based control). These studies have found that, on the one hand, indirect data-based control often performs better in the presence of \textit{variance error}, i.e., in the presence of measurement and/or process noise. On the other hand, direct data-based control typically works better in the presence of \textit{bias error}, which is the case, e.g., when the true order of the system is unknown (and hence a model of wrong order is identified) or if a linear model is used to approximate a system with nonlinear dynamics.

In this subsection, it is of our interest to make a comparison between Algorithm 1 and an indirect data-based controller for similar settings as described above. In this case, the indirect data-based control approach consists of using the collected PE data to identify the system, and then using the identified model to determine a model-based solution of the LQR problem. In particular, the approximated matrices $\hat A$ and $\hat B$ are obtained using least squares on the collected data, and then the optimal LQR controller is obtained using the command \textit{idare} in Matlab. Although more sophisticated methods for indirect data-based control exist, this simple procedure provides a good balance between accuracy and computational complexity. Besides these two methods, we also test the performance with noisy measurements of the LMI algorithm discussed in the previous subsection.

Thus, we run again 100 simulations for different linear systems with $n=5$ and $m=2$. As before, the matrices $A$ and $B$ are generated randomly with entries ranging from $-1$ to $1$. Moreover, the initial conditions $x_0$ for the persistently excited trajectory are selected randomly with entries between $-10$ and $10$. After collecting the state samples $x_k$, random uniform noise is added to form the noisy measurements $\chi_k=x_k + w_k$, with $| w_k | \leq w_{\max}$ for different values of $w_{\max}$. Table \ref{tab3} and Table \ref{tab4} show the results for Algorithm 1 (denoted as QL), the indirect data-based control method (ID), and the LMI algorithm in \cite{PeTe:21} (LMI). We observe that the ID method yields more accurate results. Although ID takes roughly twice the time compared to the QL algorithm in these experiments, the computational time is in the same order of magnitude.

We now consider the case of bias error, on which the data is collected from the nonlinear system
\begin{equation}
	x_{k+1} = Ax_k + Bu_k + 0.001 x_k^2
	\label{nlsys}
\end{equation}
where the notation $x_k^2$ represents the vector where each entry is the corresponding entry of $x_k$ squared. Notice that only a small nonlinearity affects the system (\ref{nlsys}), which otherwise behaves almost linearly. For the indirect approach, lack of model knowledge leads us to use data from (\ref{nlsys}) to identify a linear model. For the direct data-based approach, Algorithm 1 is executed using also data measured from (\ref{nlsys}).

Table \ref{tab5} displays the performance comparison between the methods QL and ID, for both the average error and the average run-time. In this case, we observe that the QL algorithm provides more accurate results while taking less time to converge. The fact that a direct data-based control method performs better in the presence of bias error is consistent with the results reported in \cite{DoCoMa:22,KrPa:21}. A comprehensive (theoretical) analysis of this finding in the considered setting of learning-based solutions to the LQR problem is an interesting issue for future research.
\begin{table}
	\centering
	\begin{tabular}{|c|c c c|} 
		\hline
		\multirow{2}{1.2cm}{\centering Magnitude of $w_{\max}$} & \multicolumn{3}{c|}{Average error $\varepsilon$} \\
		& QL & ID & LMI \\[0.5ex]
		\hline
		$0.001$ & $0.0153$ & $0.0013$ & $0.8832$ \\ 
		$0.01$ & $0.2295$ & $0.0164$ & $0.9655$ \\
		$0.1$ & $0.7091$ & $0.4011$ & $1.0648$ \\ [1ex]
		\hline
	\end{tabular}
\vspace{4pt}
	\caption{Error comparison between Algorithm 1 (QL), a model-based LQR solution (ID) and the algorithm in \cite{PeTe:21} (LMI).}
	\label{tab3}
\end{table}

\begin{table}
	\centering
	\begin{tabular}{|c|c c c|} 
		\hline
		\multirow{2}{1.2cm}{\centering Magnitude of $w_{\max}$} & \multicolumn{3}{c|}{Average run-time $(s)$} \\
		& QL & ID & LMI \\[0.5ex]
		\hline
		$0.001$ & $0.0022$ & $0.0040$ & $0.4694$ \\ 
		$0.01$ & $0.0022$ & $0.0040$ & $0.5346$ \\
		$0.1$ & $0.0023$ & $0.0041$ & $0.4878$ \\ [1ex]
		\hline
	\end{tabular}
\vspace{4pt}
	\caption{Run-time comparison between Algorithm 1 (QL), a model-based LQR solution (ID) and the algorithm in \cite{PeTe:21} (LMI).}
	\label{tab4}
\end{table}

\begin{table}
	\centering
	\begin{tabular}{|c c| c c|} 
		\hline
		\multicolumn{2}{|c|}{Average error $\varepsilon$} & \multicolumn{2}{c|}{Average time $(s)$} \\
		QL & ID & QL & ID \\[0.5ex]
		\hline
		$0.4302$ & $0.9054$ & $0.0027$ & $0.0045$ \\ [1ex]
		\hline
	\end{tabular}
\vspace{4pt}
	\caption{Comparison between Algorithm 1 (QL) and a linear model-based LQR solution (ID) for system (\ref{nlsys}).}
	\label{tab5}
\end{table}

\section{Conclusion}
\label{secconc}
An off-policy Q-learning algorithm was proposed, analyzed and compared with existing data-based control methods. The main advantage of the proposed algorithm is the reduction in computational complexity when compared, in particular, with the control-design methods based on the solutions of linear matrix inequalities. We addressed the main disadvantage of our method, namely the need of an initial stabilizing controller, by using an existing data-based deadbeat control design that does not use LMIs. Our algorithm was also shown to be robust against small disturbances in the measured data. These characteristics make the proposed algorithm an attractive alternative for practical applications of data-based LQR control.





\section*{Acknowledgment}

The authors would like to thank Prof. Tobias Damm for our fruitful discussions about the first version of this manuscript.

\ifCLASSOPTIONcaptionsoff
  \newpage
\fi




\bibliographystyle{IEEEtran}
\bibliography{IEEEabrv,learn_refs}



\end{document}